\documentclass[12pt,a4paper]{article}
\usepackage[latin9]{inputenc}
\usepackage{a4wide}
\usepackage{amssymb}
\usepackage{slashed}
\usepackage{ifpdf}
\ifpdf
\usepackage[pdftex]{graphicx}
\usepackage[pdftex,unicode,implicit]{hyperref}
\hypersetup{%
  pdftitle    = {Supersymmetric N=2 Einstein-Yang-Mills Monopoles and Covariant Attractors},
  pdfkeywords = {N=2 Supersymmetry, monopoles, black holes, attractor mechanism, BPS solutions},
  pdfauthor   = {M. Hübscher, P. Meessen, T. Ortín, S. Vaulà},
  pdfcreator  = {pdf\LaTeXe\ with package \flqq hyperref\frqq},
  pdfproducer = {pdf\LaTeXe\ with package \flqq hyperref\frqq},
  pdfpagemode = None,  
  pdffitwindow= true,  
  unicode     = true,
  plainpages  = false,
  colorlinks  = true,  
  citecolor   = blue,
  urlcolor    = red,
  linkcolor   = black
}
\newcommand{\hepth}[1]{({\tt \href{http://www.arXiv.org/abs/hep-th/#1}{hep-th/#1}})}

\else
  \usepackage[dvips]{graphicx}
  \usepackage[unicode,implicit]{hyperref}
  \newcommand{\hepth}[1]{{\tt hep-th/#1}}

\fi
\makeatletter
\@addtoreset{equation}{section}
\makeatother

\begin{document}
\begin{flushright}
\small
IFT-UAM/CSIC-07-63\\
{\tt\today}
\normalsize
\end{flushright}
\begin{center}
\vspace{1.5cm}
{\LARGE\bf Supersymmetric $N=2$ Einstein-Yang-Mills\\[.2cm] Monopoles and Covariant Attractors}
\vspace{2cm}

{\sl\large Mechthild H\"ubscher}
\footnote{E-mail: {\tt Mechthild.Huebscher@uam.es}},
{\sl\large Patrick Meessen}
\footnote{E-mail: {\tt Patrick.Meessen@uam.es}},
{\sl\large Tom{\'a}s Ort\'{\i}n}
\footnote{E-mail: {\tt Tomas.Ortin@cern.ch}}
{\sl\large and Silvia Vaul\`a}
\footnote{E-mail: {\tt Silvia.Vaula@uam.es}}

\vspace{.5cm}

{\it Instituto de F\'{\i}sica Te\'orica UAM/CSIC\\
Facultad de Ciencias C-XVI,  C.U.~Cantoblanco,  E-28049-Madrid, Spain}\\

\vspace{2cm}


{\bf Abstract}

\end{center}

\begin{quotation}
\small
We present two generic classes of supersymmetric solutions of $N=2,d=4$
supergravity coupled to non-Abelian vector supermultiplets with a gauge group
that includes an $SU(2)$ factor. The first class consists of embeddings of the 't
Hooft-Polyakov monopole and in the examples considered it has a fully regular,
asymptotically flat space-time metric without event horizons.  The other class
of solutions consists of regular non-Abelian extreme black holes. There is a
\textit{covariant attractor} at the horizon of these non-Abelian black holes.
\end{quotation}

\pagestyle{plain}


\section*{Introduction}

The search for and study of supersymmetric supergravity solutions having
the interpretation of long-range fields of string states, has been one of
the most fruitful fields of theoretical research for the last fifteen years.
In 4-dimensional theories (in particular, in $N=2$ supergravities), most of
the effort has been directed to find and study black holes with Abelian
charges in flat spacetime. The most general black-hole-type solutions of these
theories (ungauged $N=2,d=4$ supergravity coupled to vector supermultiplets)
were found in Ref.~\cite{Behrndt:1997ny}\footnote{The proof that they are the
  most general solutions of that kind was given in
  Ref.~\cite{Meessen:2006tu}, where all the supersymmetric solutions of these
  theories were found. In presence of $R^{2}$ corrections it was shown in
  Ref.~\cite{LopesCardoso:2000qm} that they have the same form as those in
  Ref.~\cite{Behrndt:1997ny}.}. This, and the existence of the attractor
mechanism \cite{attract} and its relations to stringy black hole entropy
calculations or to topological strings are two of the main results obtained so
far.

These results have not been extended to black holes with non-Abelian charges.
Actually, the little work that has been done so far in supergravity theories with
non-Abelian Yang-Mills fields concerns magnetic monopoles and not black holes.
Two main results in this direction have been the construction of two
supersymmetric gravitating monopole solutions in $N=4,d=4$ theories by Harvey
and Liu \cite{Harvey:1991jr} and Chamseddine and Volkov
\cite{Chamseddine:1997nm} whose metrics have neither singularities nor event
horizons. They have not been related to black holes and, to the best of our
knowledge there is no microscopic interpretation of these massive, charged,
objects that are not black holes but may be elementary constituents of them.
On the other hand, apart from those which are just embeddings of Abelian
solutions into non-Abelian theories, not many black-hole solutions of these
theories are known in analytic form \cite{Volkov:1998cc} and without an 
analytical form it is very
difficult to address questions about the existence of
attractors in black holes with non-Abelian charges.

Our aim in this article is to start filling this gap in our knowledge of
supersymmetric supergravity solutions with non-Abelian Yang-Mills
fields, studying, in particular, black-hole and monopole-type solutions.  We
are going to present an extension of the results of
\cite{Meessen:2006tu,Huebscher:2006mr}, characterizing the most general static
supersymmetric solutions in $N=2$ $d=4$ supergravity coupled to non-Abelian
vector supermultiplets\footnote{The general problem will be considered in
  Ref.~\cite{kn:HMOV}.}, to which we shall refer as $N=2$ $d=4$
Einstein-Yang-Mills theory. In this theory only the isometries of the
special-K\"ahler manifold parametrized by the scalars in the vector multiplets
are gauged, which leads to a positive-semidefinite scalar potential. This
characterization simplifies the search for supersymmetric black-hole solutions
and we are going to use it to study the construction of solutions in
models that admit an $SO(3)$ gauge group. We are going to present some
complete analytic solutions for two models addressing the questions
concerning attractors raised above. We are going to see that both models (and
presumably all models, including the {\em stringy} ones) admit solutions in which the
Yang-Mills fields describe an 't Hooft-Polyakov monopole and whose,
asymptotically flat, metrics are completely regular and have no event horizons,
just as it happens in the Harvey-Liu and Chamseddine-Volkov solutions. We will
also show that these models (and, again, probably all other models) admit
solutions with non-Abelian Yang-Mills fields with the same asymptotic behaviour
as the 't Hooft-Polyakov monopoles, whose metrics are regular outside an
event horizon. We will also describe how the attractor mechanism works in
these examples.
 
Monopoles in $N=2$ gauge theories were first studied by D'Adda {\em et al.\/}
in Ref.~\cite{D'Adda:1978mu}, and one of the models we are going to study is probably
its closest supergravity analogue: $SO(3)$ gauged model on $\overline{\mathbb{CP}}^{3}$.
In fact, one can see that the rigid limit of the model, see {\em e.g.\/}
\cite{Andrianopoli:1996vr}, explicitly leads to the theory studied in
\cite{D'Adda:1978mu}.  $SO(3)$ monopoles in EYM were also studied in
Ref.~\cite{Gibbons:1993xt}, but the model used there is by itself not a supergravity
theory as their scalar manifold is not special K\"ahler. 
For a specific value of the dilaton coupling \cite{Gibbons:1995zt}, however, the model corresponds to a
truncation of a supergravity and the monopoles correspond to the one found by
Harvey and Liu \cite{Harvey:1991jr}. The second model that we shall consider
closely resembles Harvey and Liu's set up and is the $\mathcal{ST}[2,n]$ model.

The plan of this article is as follows: in Section~\ref{sec-GaugedN2} we will
review $N=2$ $d=4$ supergravity coupled to non-Abelian vector supermultiplets
and describe the characterization of the supersymmetric solutions in the timelike
class, thus obtaining the minimal set of equations that need to be solved in order
to have supersymmetric solutions.  In Section~\ref{sec-solutions}, we will
discuss how one can construct solutions for generic models with an $SO(3)$
gauge group and on Sections~\ref{sec-CPn} and \ref{sec-STmn} we will consider
two specific examples of $N=2$ theory with that gauge group and we will study
the complete solutions constructed with the above methods.  Finally, in
Section~\ref{sec-concl} we will discuss our results and present our conclusions
and future directions of research, some of which are under investigation.


\section{$N=2,d=4$ EYM supergravity}
\label{sec-GaugedN2}

We start by describing the theory of $N=2$ $d=4$ supergravity coupled to
non-Abelian vector supermultiplets to which we will refer to as $N=2$
Einstein-Yang-Mills (EYM). These theories can be obtained from the ungauged
theory with vector supermultiplets by gauging the isometries of the
special-K\"ahler manifold parametrized by the scalars in the vector
supermultiplets\footnote{For a more detailed description see
  Refs.~\cite{kn:HMOV} or \cite{Andrianopoli:1996cm}, the review
  Ref.~\cite{kn:toinereview}, and the original works
  Refs.~\cite{deWit:1984pk,deWit:1984px}. Our conventions are contained in
  Refs.~\cite{Meessen:2006tu,Huebscher:2006mr}.}. We, however, do not consider
the most general gaugings, but rather restrict ourselves to gaugings that act
block-diagonally on the symplectic sections defining the theory.  Another way
of stating this property is that, in case a prepotential exists, we are going
to gauge symmetries of the prepotential. Hence, we are going to gauge groups
$G\subseteq Sl(\overline{n},\mathbb{R})$, where $n=\bar{n}-1$ is the number
of vector multiplets.

The bosonic part of the action for these theories is given by

\begin{equation}
\label{eq:action}
\begin{array}{rcl}
 S & = & {\displaystyle\int} d^{4}x \sqrt{|g|}
\left[R +2\mathcal{G}_{ij^{*}}\mathfrak{D}_{\mu}Z^{i}
\mathfrak{D}^{\mu}Z^{*\, j^{*}}
+2\Im{\rm m}\mathcal{N}_{\Lambda\Sigma} 
F^{\Lambda\, \mu\nu}F^{\Sigma}{}_{\mu\nu}
 \right. \\
& & \\
& & \left. 
\hspace{2cm}
-2\Re{\rm e}\mathcal{N}_{\Lambda\Sigma}  
F^{\Lambda\, \mu\nu}{}^{\star}F^{\Sigma}{}_{\mu\nu}
-V(Z,Z^{*})
\right]\, ,
\end{array}
\end{equation}

\noindent
and, for vanishing fermions, the supersymmetry transformation rules of the
fermions are

\begin{eqnarray}
\delta_{\epsilon}\psi_{I\, \mu} & = & 
\mathfrak{D}_{\mu}\epsilon_{I} 
+\epsilon_{IJ}T^{+}{}_{\mu\nu}\gamma^{\nu}\epsilon^{J}\, ,
\label{eq:gravisusyrule}\\
& & \nonumber \\
\delta_{\epsilon}\lambda^{Ii} & = & 
i\not\!\!\mathfrak{D} Z^{i}\epsilon^{I} 
+\epsilon^{IJ}[\not\!G^{i\, +} +W^{i}]\epsilon_{J}\, .
\label{eq:gaugsusyrule}
\end{eqnarray}

\noindent
The supersymmetry transformation laws for the bosons are the same as in the
ungauged case \cite{Meessen:2006tu}.  This will have important consequences in
what follows.

In the above equations

\begin{equation}
\label{eq:nablazio}
\mathfrak{D}_{\mu} Z^{i} = \partial_{\mu} Z^{i}+gA^{\Lambda}{}_{\mu} 
k_{\Lambda}{}^{i}\, , 
\end{equation}

\noindent
where $k_{\Lambda}{}^{i}(Z)$ are the holomorphic Killing vectors of the metric
$\mathcal{G}_{ij^{*}}$, is the gauge covariant derivative acting on the
scalars; the covariant derivative acting on the Killing spinor is

\begin{equation}
\label{eq:dei}
\mathfrak{D}_{\mu}\epsilon_{I} =
\left\{
\nabla_{\mu}+{\textstyle\frac{i}{2}}\mathcal{Q}_{\mu}
+{\textstyle\frac{i}{2}}gA^{\Lambda}{}_{\mu}\mathcal{P}_{\Lambda}
\right\}\epsilon_{I}\, ,
\end{equation}

\noindent
where $\mathcal{Q}_{\mu}$ is the pullback of the  K\"ahler 1-form and
$\mathcal{P}_{\Lambda}$ is the momentum map satisfying 

\begin{equation}
\label{eq:prepo}
k_{\Lambda\, i^{*}} =i\partial_{i^{*}}\mathcal{P}_{\Lambda}\, ,  
\end{equation}

\noindent
is the Lorentz- K\"ahler- and gauge-covariant derivative acting on spinors
and

\begin{equation}
W^{i}=
{\textstyle\frac{1}{2}}g\mathcal{L}^{*\, \Lambda}k_{\Lambda}{}^{i}\, .  
\end{equation}

The potential $V(Z,Z^{*})$,  is given by

\begin{equation}
V(Z,Z^{*}) \ =\ 2\mathcal{G}_{ij^{*}}W^i W^*{}^{j^*}
          \ =\ -{\textstyle\frac{1}{4}}g^{2} (\Im{\rm m}\mathcal{N})^{-1|\Lambda\Sigma}
                \mathcal{P}_{\Lambda}\mathcal{P}_{\Sigma}\, .
\end{equation}

The negative-definiteness of $\Im{\rm m}\mathcal{N}_{\Lambda\Sigma}$ and the
reality of the momentum map imply that $V\geq 0$.  

We are interested in supersymmetric solutions of the above system admitting at
least one Killing spinor $\epsilon_{I}$. Their general form can be found following
Refs.~\cite{Meessen:2006tu,Huebscher:2006mr} and our discussion will be extremely
brief. As usual, if $\epsilon_{I}$ is a Killing spinor, the
bilinear $V^{\mu}=i\bar{\epsilon}^{I}\gamma^{\mu}\epsilon_{I}$ is a
non-spacelike Killing vector. We consider only the case in which it is
timelike, i.e. $V^{2}=4|X|^{2}>0$, and introduce a time coordinate $t$ by
$V^{\mu}\partial_{\mu}=\sqrt{2}\partial_{t}$.

{}From the gaugino variation (\ref{eq:gaugsusyrule}) we get the equation

\begin{equation}
  \label{eq:VdZ}
V^{\mu}\mathfrak{D}_{\mu}Z^{i}+2X W^{i}=0\, ,
\end{equation}

\noindent
whose analogue in the ungauged case states that the scalars $Z^{i}$ are
time-independent.  In the gauged case, we can obtain time-independence by
choosing the gauge fixing

\begin{equation}
\label{eq:gaugechoice2}
   A^{\Lambda}{}_{t}  \; = \;   
       -\sqrt{2}|X|^{2}\ \mathcal{R}^{\Lambda}\, ,
\end{equation}

\noindent 
which solves Eq. (\ref{eq:VdZ}) due to the property
$\mathcal{L}^{\Lambda}k_{\Lambda}^{i}=0$ \cite{Andrianopoli:1996cm}.

The other three vectors $V^{m}\,\,\, (m=1,2,3)$ that we can construct as
bilinears of the Killing spinor are exact 1-forms, and can thence be used to define spatial
coordinates $x^{m}$ by $V^{m}\equiv dx^{m}$. The metric is of the
conformastationary form

\begin{equation}
\label{eq:metric}
ds^{2} \; =\; 2|X|^{2}(dt+\omega)^{2} 
-\frac{1}{2|X|^{2}}dx^{m}dx^{m}\, ,
\end{equation}

\noindent
where $\omega=\omega_{m}dx^{m}$ is a possible 1-form.
As we are interested in static spacetimes we are going to take $\omega =0$.
This choice imposes a constraint which can be written as

\begin{equation}
  \label{eq:29}
\langle\ \mathcal{I}\ |\ \mathfrak{D}_{m}\mathcal{I}\rangle  \; =\; 0 \; .
\end{equation}

\noindent
where we have used the variables that will govern the solutions:

\begin{equation}
\label{eq:realsections}
\mathcal{R}\equiv \Re{\rm e}\, (\mathcal{V}/X)\; ,\; 
\mathcal{I}\equiv \Im{\rm m}\, (\mathcal{V}/X)\;\;\longrightarrow\;\;
\frac{1}{2|X|^{2}} \ =\ \langle\, \mathcal{R}\mid\, \mathcal{I}\, \rangle \; ,
\end{equation}

Observe that up to the replacement of the ordinary derivative by the gauge-covariant
derivative, the constraint has the same functional form as in the Abelian
case.

{}From the gravitino and gaugino variations we deduce the symplectic vector of
2-form field-strengths:

\begin{equation}
\label{eq:FSuper}
F \; =\; -\sqrt{2}
\ \mathfrak{D}\left(\ |X|^{2}\mathcal{R}\ dt\right) 
\; -\sqrt{2}\; |X|^{2}\ \star \left( dt\wedge\ \mathfrak{D} \mathcal{I}\right) \, ,  
\end{equation}

\noindent
which, again, has the same functional form as in the ungauged case
and is, moreover, consistent with the gauge fixing (\ref{eq:gaugechoice2})!

{}Following the steps outlined in \cite[Sec. 4.3]{Meessen:2006tu}
one can readily check that all the configurations of the above form are, at least,
1/2-BPS.

We still have to impose the equations of motion in order to find
supersymmetric solutions.  
As any other symmetry of an action functional, supersymmetry implies relations
between equations of motion. In contrast to other symmetries, however, 
supersymmetry implies relations between the e.o.m. of fields of different
spin. This opens up the possibility to find a minimal set of e.o.m.s that
need to be solved explicitly as to insure that all e.o.m.s are solved.

The most economical way of finding such a minimal set is
by means of the Killing Spinor Identities (KSIs)
\cite{Kallosh:1993wx,Bellorin:2005hy}, which are off-shell relations between
the equations of motion of the bosons of a supersymmetric theory. 
A remarkable characteristic of the KSIs is that their functional form
depends only on the structure of the supersymmetry transformation rules of the bosons.
As we remarked above, in the case of $N=2$ EYM the supersymmetry transformations 
of the bosons are exactly the same as in the
ungauged case: this implies that the KSIs have the same form as in the ungauged case (given in Ref.~\cite[Sec.
3.1.1]{Meessen:2006tu}) even if the equations of motion are different. 
Seeing this, we must conclude that in order to be sure that the configuration
we obtained above solves the equations of motions, we only need to impose the 
Bianchi identities and the Yang-Mills equation.

The Bianchi identities for the supersymmetric field strengths
Eq.~(\ref{eq:FSuper}) take the form

\begin{equation}
  \label{eq:Maxwell9}
  \mathfrak{D}_{m}\mathfrak{D}_{m}\ \mathcal{I}^{\Lambda} \; =\; 0 \; ,
\end{equation}

\noindent
and the YM equations take the form


\begin{equation}
  \label{eq:Maxwell16}
  \mathfrak{D}_{m}\mathfrak{D}_{m}\mathcal{I}_{\Lambda} \; =\; 
      \textstyle{1\over 2}g^{2}\ \left[
          f_{\Lambda (\Sigma}{}^{\Gamma}f_{\Delta )\Gamma}{}^{\Omega}\ \mathcal{I}^{\Sigma}\mathcal{I}^{\Delta}
      \right]\; \mathcal{I}_{\Omega} \; .
\end{equation}

Eqs.~(\ref{eq:Maxwell9}) and (\ref{eq:Maxwell16}) form a complicated system to
solve, but not as complicated as one might have anticipated: in principle one
might have imagined the appearance of $\mathcal{R}$ in the system which,
seeing that they are functions of $\mathcal{I}$, would make the solutions
highly non-linear. Fortunately this does not happen and we end up with a nice hierarchical picture:
first solve (\ref{eq:Maxwell9}) as to obtain the pair
$(A^{\Lambda},\mathcal{I}^{\Lambda})$ and use this information to find a
solution to Eq.~(\ref{eq:Maxwell16}).  Then solve the stabilization equations
to obtain $\mathcal{R}$ and use this to calculate $|X|^{2}$ through Eq.
(\ref{eq:realsections}). Lastly, as we imposed staticity, we must check Eq.
(\ref{eq:29}) and its integrability equation

\begin{equation}
\label{eq:30}
\langle\ \mathcal{I}\ |\ \mathfrak{D}_{m}\mathfrak{D}_{m}\mathcal{I}\rangle
\; =\; 0 \; ,
\end{equation}

\noindent
in order to avoid singularities like the ones studied in
Refs.~\cite{Denef:2000nb,Bellorin:2006xr}.

Eq.~(\ref{eq:Maxwell9}) is, of course, a hard nut to crack, and it is 
a better idea to start with a given $A^{\Lambda}$ and try to distill
an $\mathcal{I}^{\Lambda}$ from it by comparing the resulting
field strength with the expression (\ref{eq:FSuper}). 
Doing so, we find Eq.~(\ref{eq:gaugechoice2}) and

\begin{equation}
\label{eq:28}
\textstyle{1\over 2}\ \epsilon_{pmn}\ F^{\Lambda}_{mn} \; =\; 
-\textstyle{1\over\sqrt{2}}\ \mathfrak{D}_{p}\mathcal{I}^{\Lambda} \; .
\end{equation}

\noindent
This equation is readily recognised as the Bogomol'nyi equation
\cite{kn:Bog} and allows us to embed YM solutions satisfying it
(\textit{e.g.}~monopoles) into $N=2$ EYM theories. In the next section we
are going to work out some of these solutions.


\section{Solutions of $SO(3)$ $N=2$ EYM}
\label{sec-solutions}

For brevity let us only consider $N=2$ EYM systems containing an $SO(3)$ gauge
group, parametrizing the directions in which the $SO(3)$ acts with indices
$a=1,2,3$ and ignoring for the moment the other directions. If we make the ``hedgehog''
Ansatz

\begin{equation}
  \label{eq:33}
  \mathcal{I}^{a} \; =\; \mathcal{I}(r)\ n^{a}\, ,
\hspace{1cm}
  A^{a}{}_{m} \; =\; \Phi(r)\ \varepsilon_{mn}{}^{a}\ n^{n} \; ,
\hspace{1cm}
n^{a}\equiv x^{a}/r\, ,
\hspace{1cm}
r\equiv \sqrt{x^{b}x^{b}}\, .
\end{equation}

\noindent
where $\mathcal{I}$ and $\Phi$ are functions of $r$ alone, we see that the
Bogomoln'nyi equation (\ref{eq:28}) admits a 2-parameter ($\mu$ and $\rho$)
family of solutions given by \cite{Protogenov:1977tq}

\begin{equation}
\label{eq:43}
\begin{array}{rclrcl}
\mathcal{I}(r) & = &  
{\displaystyle \frac{\sqrt{2}\mu}{g}}\mathsf{H}_{\rho}(\mu r)\, ,
& 
\mathsf{H}_{\rho}(r) & = & {
\displaystyle\coth{(r+\rho)} \ -\ \frac{1}{r}} \; ,  
\\
& & \\
\Phi(r) & = & 
{\displaystyle \frac{\mu}{g}} \mathsf{G}_{\rho}(\mu r)\, ,
& 
\mathsf{G}_{\rho}(r) & = & 
{\displaystyle\frac{1}{r} \ -\ \frac{1}{\sinh{(r+\rho)}}}\, . \\
\\
\end{array}
\end{equation}

The next step is to obtain the $\mathcal{I}_{a}$ from
Eq.~(\ref{eq:Maxwell16}): a solution to this equation is readily found by
observing that since $\mathcal{I}_{a}$ has to be proportional to $n^{a}$, the r.h.s.
of said equation vanishes identically. Also, the co-adjoint representation
under which $\mathcal{I}_{a}$ transforms, is the same as the adjoint
representation, whence Eq.~(\ref{eq:Maxwell16}) reduces to Eq.
(\ref{eq:Maxwell9}). The result is

\begin{equation}
  \label{eq:38}
  \mathcal{I}_{a} \; =\; \frac{g\mathcal{J}}{2}\  \mathcal{I}^{a} \; ,
\end{equation}

\noindent
where $\mathcal{J}$ is an arbitrary constant. 

The fact that $\mathcal{I}_{a}$ has the same functional form as 
$\mathcal{I}^{a}$ has consequences for the staticity condition
Eq.~(\ref{eq:29}): if we split the index $\Lambda$ into an $a$-index and an
$u$-index labelling the $u$ngauged directions, we see that the condition
(\ref{eq:29}) acts non-trivially only on the ungauged part, {\em i.e.}

\begin{equation}
\label{eq:41}
 \mathcal{I}_{u}\ d\mathcal{I}^{u} \ 
-\ \mathcal{I}^{u}\ d\mathcal{I}_{u}\ 
+\ \mathcal{I}_{a}\ \mathfrak{D}\mathcal{I}^{a} \ 
-\ \mathcal{I}^{a}\ \mathfrak{D}\mathcal{I}_{a}
; =\; \mathcal{I}_{u}\ d\mathcal{I}^{u} \ 
-\ \mathcal{I}^{u}\ d\mathcal{I}_{u} \; =\; 0 \; ,
\end{equation}

\noindent 
which we can therefore solve as in the Abelian case. 

At this point the solutions are completely determined. 
In order to find the explicit
forms of $\mathcal{R}$ and the spacetime metric, however, we must solve the
stabilization equations which depend on the specific model under consideration. We will
study two models that allow for an $SO(3)$ gauging in Sections~\ref{sec-CPn} and
\ref{sec-STmn} and there we will discuss the physical properties of the
complete solutions.

For now we are going to study two particularly interesting solutions of
the above family: those with $\rho=0$ and those with $\rho\rightarrow\infty$.


\subsection{$\rho=0$: 't Hooft-Polyakov Monopoles}

The $\rho=0$ solution can be written, in our normalization, in the form

\begin{equation}
\label{eq:37}
\begin{array}{rclrcl}
A^{a}{}_{m} & =& \varepsilon_{mb}{}^{a}\ n^{b}\ 
{\displaystyle\frac{\mu}{g}\ \mathsf{G}_{0}(\mu r)}\, ,\hspace{1cm} &
\mathsf{G}_{0}(r) & =& {\displaystyle\frac{1}{r} \ -\ \frac{1}{\sinh r}} \; , \\
&& && & \\
\mathcal{I}^{a} & = & 
{\displaystyle\frac{\sqrt{2}\mu}{g}\ \mathsf{H}_{0}(\mu r)\ n^{a}}\, , &
\mathsf{H}_{0}(r) & =& {\displaystyle\coth r \ -\ \frac{1}{r}} \; ,\\  
&& && & \\
\mathcal{I}_{a} & = & 
{\displaystyle\frac{\mu\mathcal{J}}{\sqrt{2}}\ 
\mathsf{H}_{0}(\mu r)\ n^{a}}\, . &
& & \\
\end{array}
\end{equation}

The profile of the functions $\mathsf{G}_{0}$ and $\mathsf{H}_{0}$ are given
Fig.~(1). 
These functions are regular and bound between $0$ and
$1$ and . Thus, we see that $\mathcal{I}$ (whence also $\mathcal{I}^{a}$ and
$\mathcal{I}_{a}$) are regular at $r=0$. The YM fields of this
solution are those of the 't Hooft-Polyakov monopole \cite{'t
  Hooft:1974qc}.

\begin{figure}
\label{fig:profiles}
\centering
\includegraphics[width=8cm]{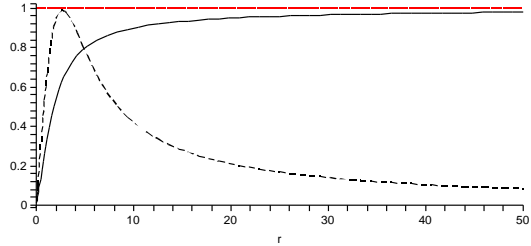}
\caption{The profiles of the functions $\mathsf{G}_{0}$ and $\mathsf{H}_{0}$.}
\end{figure}


\subsection{$\rho\rightarrow \infty$: Black hedgehogs}

In the limit $\rho\rightarrow\infty$ the solution becomes

\begin{equation}
\label{eq:35}
\begin{array}{rcl}
A^{a}{}_{m} &  = &  \varepsilon_{mb}{}^{a}\ {\displaystyle\frac{n^{b}}{gr}}\; ,\\
& & \\
\mathcal{I}^{a} & = & -\sqrt{2}\left( \mathcal{I}_{\infty} \ 
+\ {\displaystyle\frac{1}{gr}}\right) \, n^{a} \; ,
\hspace{1cm}
 \mathcal{I}_{\infty} \equiv -{\displaystyle\frac{\mu}{g}}\, ,
\\
& & \\
\mathcal{I}_{a} & = & -{\displaystyle\frac{g\mathcal{J}}{\sqrt{2}}}
\left( \mathcal{I}_{\infty} \ 
+\ {\displaystyle\frac{1}{gr}}\right) \, n^{a} \; .\\
 \end{array}
\end{equation}

These fields are singular at $r=0$.  This singularity makes the solution
uninteresting in flat spacetime and this is, probably, the reason why it has
not been considered before in the literature. However, the coupling to gravity
may cover it by an event horizon in which case we would obtain a non-Abelian
black hole solution which we call a ``black hedgehog''.


\section{Embedding in $\overline{\mathbb{CP}}^{n}$ models}
\label{sec-CPn}

As mentioned in Section~\ref{sec-GaugedN2}, in order to construct models of
gauged sugra one can start with the invariances of a prepotential
$\mathcal{F}$. Some of the easiest models are those given by quadratic
prepotentials and in the case of $\overline{\mathbb{CP}}^{n}$ the prepotential
reads\footnote{Even though we will fix $n=3$ in following subsections, for the
  moment we will leave $n$ undetermined. }

\begin{equation}
\label{eq:30a}
\mathcal{F} \; =\; 
\textstyle{i\over 4}\ \eta_{\Lambda\Sigma}\ \mathcal{X}^{\Lambda}\
\mathcal{X}^{\Sigma}\, , 
\hspace{.5cm}
\eta \ =\ \mathrm{diag}\left(\ -\ ,\ [+]^{n}\ \right) \; .
\end{equation}




\noindent
The K\"ahler potential is straightforwardly derived to give 

\begin{equation}
\label{eq:31}
e^{-\mathcal{K}} \; =\; 
|\mathcal{X}^{0}|^{2} \ -\ \sum_{i=1}^{n}\ |\mathcal{X}^{i}|^{2}
\; =\; 1\ -\ \sum_{i=1}^{n}\ |Z^{i}|^{2} \equiv 
 1\ -\ |Z|^{2} \; ,
\end{equation}

\noindent
resulting in the 
standard Fubini-Study metric on $\overline{\mathbb{CP}}^{n}$

\begin{equation}
\label{eq:31a}
\mathcal{G}_{ij^{*}} \; =\; 
\frac{\delta_{ij^{*}}}{1\ -\ |Z|^{2}} \; +\; 
\frac{Z^{i}\ Z^{*j^{*}} }{\left( 1\ -\ |Z|^{2} \right)^{2}}\, , 
\hspace{.5cm}
\mathcal{G}^{ij^{*}} \; =\; 
\left( 1 -|Z|^{2}\right)\ 
\left\{ \delta^{ij^{*}}\ -\ Z^{i}Z^{*j^{*}}  \right\}\; ,
\end{equation}

\noindent
which is an Einstein metric, {\em i.e.\/} $R(\mathcal{G}) =\ \bar{n}\ \mathcal{G}$. In
fact, $\overline{\mathbb{CP}}^{n}$ can be identified with the symmetric space
$SU(1,n)/U(n)$. Observe that Eq.~(\ref{eq:31}) the coordinates $Z^{i}$ are
constrained by

\begin{equation}
\label{eq:constraint}
0 \leq |Z|^{2} < 1\, .  
\end{equation}

The stabilization equations can be readily solved in this model:

\begin{equation}
\label{eq:9}
\mathcal{R}_{\Lambda} \; =\; 
-\textstyle{1\over 2}\eta_{\Lambda\Sigma}\ \mathcal{I}^{\Sigma} \;\;\;\; ,\;\;\;\;
\mathcal{R}^{\Lambda} \; =\; 2\eta^{\Lambda\Sigma}\ \mathcal{I}_{\Sigma} \; ,
\end{equation}

\noindent
which allows us to write down the metrical factor in 
Eq.~(\ref{eq:realsections}) in terms of the $\mathcal{I}^{\Lambda}$ and
$\mathcal{I}_{\Lambda}$ as

\begin{equation}
\label{eq:17}
-g_{rr}= \frac{1}{2|X|^{2}} =
-\textstyle{1\over 2}\ 
\mathcal{I}^{\Lambda}\eta_{\Lambda\Sigma}\mathcal{I}^{\Sigma}
\ -\ 2\ \mathcal{I}_{\Lambda}\eta^{\Lambda\Sigma}\mathcal{I}_{\Sigma} \; =\;
\textstyle{1\over 2}
\left[ \mathcal{I}^{0}{}^{2} \ -\ \mathcal{I}^{i}{}^{2} 
\ +\ 4\mathcal{I}_{0}^{2} \ -\ 4\mathcal{I}_{i}^{2}
\right] \; .
\end{equation}

Let us then consider the case $n=3$: due to Eq.~(\ref{eq:38}) $\mathcal{I}_{a}$ is proportional
to $\mathcal{I}^{a}$ and using the hedgehog
Ansatz Eq.~(\ref{eq:33},\ref{eq:43}) we obtain

\begin{equation}
-g_{rr}=
\frac{1}{2|X|^{2}} =
\textstyle{1\over 2}
\left\{ \mathcal{I}^{0}{}^{2} 
\ +\ 4\mathcal{I}_{0}^{2} 
\ -\ 2\mu^{2}\left[{\displaystyle\frac{1}{g^{2}}}+\mathcal{J}^{2}\right]
\mathsf{H}^{2}_{\rho}(\mu r) 
\right\}\; .
\end{equation}

At the level of the metric the system behaves as if
we were dealing with a $U(1)$ field instead of an $SU(2)$ field!

Let us then try to find a regular embedding of the 't Hooft-Polyakov
monopole in the $\overline{\mathbb{CP}}^{3}$ model: since the function
$\mathsf{H}_{0}(\mu r)$ is bound, it is enough for $\mathcal{I}^{0}$ and
$\mathcal{I}_{0}$ to be constant as to insure that the scalars satisfy the
constraint Eq.~(\ref{eq:constraint}). Actually, taking them to be spherically
symmetric, harmonic functions would produce scalars that violate said constraint
and introduce singularities. Fixing the values of $\mathcal{I}^{0}$ and
$\mathcal{I}_{0}$ by imposing asymptotic flatness we find

\begin{equation}
\label{eq:39}
-g_{rr}=\frac{1}{2|X|^{2}}= 
1 \ +\ \mu^{2}\left[ \frac{1}{g^{2}} \ +\ \mathcal{J}^{2}\right]\
\left( 1\ -\ \mathsf{H}^{2}(\mu r)\right) \; ,
\end{equation}

\noindent
which means that the metric is perfectly regular and describes an object of
mass

\begin{equation}
\label{eq:40}
\mathsf{M} \; =\; \mu\left[  \frac{1}{g^{2}} \ +\ \mathcal{J}^{2}\right] \; .
\end{equation}

Let us now consider the black hedgehog case. Since the function
$\mathsf{H}_{\infty}(\mu r)$ is singular, in order to produce scalar fields
that satisfy the bound Eq.~(\ref{eq:constraint}), either $\mathcal{I}^{0}$ or
$\mathcal{I}_{0}$ has to be unfrozen, {\em i.e.\/} a non-constant harmonic function.

Choosing for simplicity

\begin{equation}
\mathcal{I}^{0} = \mathcal{I}^{0}_{\infty} + \frac{p^{0}}{r}\, ,
\end{equation}

\noindent
we get

\begin{equation}
  \begin{array}{rcl}
-g_{rr}={\displaystyle\frac{1}{2|X|^{2}}} & = & 
\textstyle{1\over 2}
\left\{ \mathcal{I}_{\infty}^{0}{}^{2} -2\mu^{2}
\left[  {\displaystyle\frac{1}{g^{2}}} 
\ +\ \mathcal{J}^{2}\right]
\right\} \\
& & \\
& & 
+
\left\{ \mathcal{I}_{\infty}^{0}p^{0} -2|\mu|
\left[  {\displaystyle\frac{1}{g^{2}}} 
\ +\ \mathcal{J}^{2}\right]
\right\}{\displaystyle\frac{1}{r}} \\
& & \\
& & 
+
\textstyle{1\over 2}
\left\{ p^{0}{}^{2} -2
\left[  {\displaystyle\frac{1}{g^{2}}} 
\ +\ \mathcal{J}^{2}\right]
\right\} {\displaystyle\frac{1}{r^{2}}}
\; .\\
\end{array}
\end{equation}

The first term has to be normalized to 1 to have asymptotic flatness. The
coefficient of the second term is the mass and should be positive; the
coefficient of the last term, iff positive, is the area of an event
horizon divided by $4\pi$. A metric describes a regular
black hole if it is asymptotically flat, has a horizon and the mass and the entropy are
positive definite. 

It is always possible to choose the parameters such as
to obtain a regular black hole. A simple choice is

\begin{equation}
 \mathcal{I}_{\infty}^{0} = \sqrt{2}\sqrt{1+\mu^{2}
\left[  {\displaystyle\frac{1}{g^{2}}}
\ +\ \mathcal{J}^{2}\right]}
\, ,
\hspace{1cm}
p^{0} = |\mu|^{-1}  \mathcal{I}_{\infty}^{0}\, ,
\end{equation}

\noindent
and gives a mass and event horizon area

\begin{equation}
\begin{array}{rcl}
\mathsf{M} & =& 2 |\mu|^{-1}\, ,\\
& & \\
\mathsf{A} & = & 4\pi |\mu|^{-2}\, .\\  
\end{array}
\end{equation}

On the event horizon the scalars $\mathcal{Z}^{a}$ take the values

\begin{equation}
Z^{a} = \frac{\sqrt{2}}{p^{0}}\, \left(\frac{1}{g}-i\mathcal{J}\right)\, n^{a}\, ,  
\end{equation}

\noindent
which are independent of their asymptotic values, but not constant over the
horizon. Actually, since these scalars are charged, the most we can ask
for is that they are constant up to $SO(3)$ gauge transformations
(\textit{i.e.}~covariantly constant), which is the case. The scalar
fields have a \textit{covariant attractor} on the horizon and their
gauge-invariant combination $|Z|^{2}$ has a standard attractor.


\section{Embedding in $\mathcal{ST}[2,n]$ models}
\label{sec-STmn}

We are now going to consider the embedding of the 't Hooft-Polyakov monopole
and the black hedgehog into a more stringy model of the $\mathcal{ST}[2,n]$
family.

Let us start  by giving the symplectic section

\begin{equation}
\label{eq:STSection}
\Omega \; =\; 
\left( 
  \begin{array}{c}
\mathcal{X}^{\Lambda} \\ \eta_{\Lambda\Sigma} \mathcal{X}^{\Sigma}\mathcal{S}\\
\end{array}
\right) 
\;\;\; \mbox{with}\;\;\; \mathcal{X}\cdot \mathcal{X}
\equiv \eta_{\Lambda\Sigma}\mathcal{X}^{\Lambda}\mathcal{X}^{\Sigma} \ =\ 0 \; , 
\end{equation}

\noindent
where the metric $\eta = \mathrm{diag}([+]^{2},[-]^{n})$. In this
parameterization no prepotential exists but we can do a symplectic
transformation such that a prepotential exists.
The K\"ahler potential is

\begin{equation}
  \label{eq:STKahlPot}
  e^{-\mathcal{K}} \; =\; -2\ \Im{\rm m} \mathcal{S}\ 
\mathcal{X}\cdot \mathcal{X}^{*}\; .
\end{equation}

The stabilisation equation was solved in Ref.~\cite{Kallosh:1996tf}. Using the
notation

\begin{equation}
p^{\Lambda} \equiv \Im {\rm m}\mathcal{X}^{\Lambda}\, ,
\hspace{1cm}   
q_{\Lambda} \equiv  \eta_{\Lambda\Sigma}
\Im {\rm m}(\mathcal{S}\mathcal{X}^{\Lambda})\, ,
\end{equation}

\noindent
the solution takes the form

\begin{equation}
\label{eq:STStabSol}
\mathcal{S} =
\frac{ p\cdot q}{p\cdot p} 
\ +\ i\frac{\left( p\cdot p\ q\cdot q \ -\ (p\cdot q)^{2}   
\right)^{1/2} }{p\cdot p} \; , 
\hspace{1cm}
\mathcal{X}\cdot\mathcal{X}^{*}  = p\cdot p \; ,.
\end{equation}

Knowing the solution to the stabilisation equation it is straightforward to
derive the metrical factor of our solutions as

\begin{equation}
\label{eq:STMetFact}
-g_{rr}=\frac{1}{2|X|^{2}} \; =\; 
\sqrt{ p\cdot p\ q\cdot q \ -\ (p\cdot q)^{2} } \; ,
\end{equation}

\noindent
where we must substitute $p^{\Lambda}=\mathcal{I}^{\Lambda}$ and $q_{\Lambda}
=\mathcal{I}_{\Lambda}$.

Let us then restrict ourselves to the $\mathcal{ST}[2,3]$ model 
and gauge the $SO(3)$ group. Using
indices $i,j=1,2$ for the first two components (which we assume correspond to
ungauged directions) and taking into account Eqs.~(\ref{eq:38},\ref{eq:33})
and (\ref{eq:43}), the metric factor can be written in the form

\begin{equation}
\label{eq:STMetFact2}
-g_{rr}=\frac{1}{2|X|^{2}} \; =\; 
\sqrt{ \mathcal{I}^{i}\mathcal{I}^{i} \mathcal{I}_{j}\mathcal{I}_{j} 
-(\mathcal{I}^{i}\mathcal{I}_{i})^{2}
-\left(\mathcal{I}_{i}
-\frac{g\mathcal{J}}{2} \mathcal{I}^{i}\right)
\left(\mathcal{I}_{i}
-\frac{g\mathcal{J}}{2} \mathcal{I}^{i}\right)
\mathsf{H}_{\rho}^{2}(\mu r)
} \; .
\end{equation}

Again, the $SU(2)$ fields enter effectively the metric as a $U(1)$ field.

The 't Hooft-Polyakov monopole can be given a regular embedding in this model
by taking the $\mathcal{I}^{i}$ and  $\mathcal{I}_{i}$ to be constant. The
metric function takes the form

\begin{equation}
\label{eq:STMetFact3}
-g_{rr}=\frac{1}{2|X|^{2}} \; =\; 
\sqrt{ 1 \ +\ 2\mu\mathsf{M}\left[ 1\ -\ \mathsf{H}_{0}^{2}(\mu r)\right] }\, ,
\end{equation}

\noindent
where we have normalized

\begin{equation}
\mathcal{I}^{i}\mathcal{I}^{i} \mathcal{I}_{j}\mathcal{I}_{j} 
-(\mathcal{I}^{i}\mathcal{I}_{i})^{2}
-2\mu^{2}\left(\frac{1}{g}\mathcal{I}_{i}
-\frac{\mathcal{J}}{2} \mathcal{I}^{i}\right)
\left(\frac{1}{g}\mathcal{I}_{i}
-\frac{\mathcal{J}}{2} \mathcal{I}^{i}\right) =1\, .
\end{equation}

\noindent
The mass is given by 

\begin{equation}
\label{eq:STMmass}
\mathsf{M} =
\mu\left(\frac{1}{g}\mathcal{I}_{i}
-\frac{\mathcal{J}}{2} \mathcal{I}^{i}\right)
\left(\frac{1}{g}\mathcal{I}_{i}
-\frac{\mathcal{J}}{2} \mathcal{I}^{i}\right)\, ,
\end{equation}

\noindent
and is manifestly positive if $\mu$ is, automatically making the metric
completely regular. Again, the spacetime has no event horizons.

The black hedgehog can also be given a regular embedding in this model, and
requires of the introduction of a second unfrozen $U(1)$ field. By choosing
the parameters judiciously, the mass will be positive and the area of the horizon will be
finite, leading to a regular black hole. The comments on covariant attractors made in the
$\overline{\mathbb{CP}}^{n}$ case apply to this case without any change.


\section{Conclusions and outlook}
\label{sec-concl}

In this paper we have given the general recipe to construct supersymmetric
solutions in the timelike class of $N=2$ Super Einstein-Yang-Mills theories
and we have shown the generic existence in $N=2$ 
Einstein-Yang-Mills theories with an $SO(3)$ gauge group of regular, extreme,
supersymmetric non-Abelian black holes (black hedgehogs) and monopoles.  The
monopole solutions found long ago in
Refs.~\cite{Harvey:1991jr,Chamseddine:1997nm} should be
particular examples of this general class of monopole solutions. On the other
hand, the $SU(2)\times U(1)$ black hole solution of Ref.~\cite{Kallosh:1994wy}
should also belong to the class of black hedgehogs, although finding the exact
correspondence is a difficult task.

We have shown that, at least in the cases considered, there is a
\textit{covariant attractor mechanism} and work on a general proof in under
way.


\section*{Acknowledgements}

This work has been supported in part by the Spanish Ministry of Science and
Education grants FPU AP2004-2574 (MH), FPA2006-00783 and PR2007-0073 (TO), the Comunidad de
Madrid grant HEPHACOS P-ESP-00346, by the EU Research Training Network
\textit{Constituents, Fundamental Forces and Symmetries of the Universe}
MRTN-CT-2004-005104 and by the {\em Fondo Social Europea} through an
I3P-doctores scholarship (PM).  PM wishes to thank R.~Hern\'andez,
K.~Landsteiner, E. L\'opez and C. Pena for discussions.
TO wishes to thank the Stanford Institute for Theoretical Physics for 
their hospitality, and M.M. Fern\'andez for her lasting support.


\end{document}